\documentclass[reprint,superscriptaddress,10pt, amsmath,amssymb, aps, pra]{revtex4-2}
\usepackage[utf8]{inputenc}
\usepackage[T1]{fontenc}
\usepackage{graphicx}             
\usepackage{xcolor}
\definecolor{darkblue}{rgb}{0, 0, 0.8}
\usepackage[colorlinks=true, breaklinks=true, linkcolor=darkblue, citecolor=darkblue, urlcolor=darkblue]{hyperref}
\usepackage{braket}
\usepackage{soul} 
\usepackage{wrapfig}
\usepackage{physics}
\usepackage{ulem}
\usepackage{eqnarray}
\usepackage{stmaryrd}
\usepackage[english]{babel}

\definecolor{green2}{rgb}{0., 0.5, 0.}
\definecolor{blue2}{rgb}{0., 0.35, 1}
\definecolor{brown}{rgb}{0.7, 0.15, 0}
\definecolor{red}{rgb}{1, 0.1, 0.1}

\newcommand{\im}{\mathrm{i}}
\newcommand{\disc}{\mathrm{\disc}}

\begin{document}

\title{Simulation of Hopfield-like Hamiltonians using time-multiplexed photonic networks}

\author{T. Seck}
    \affiliation{NTT Research, Inc. Physics \& Informatics Laboratories, 940 Stewart Dr, Sunnyvale, CA 94085}%
    \affiliation{IPCMS (UMR 7504) and ISIS (UMR 7006), University of Strasbourg and CNRS, 67000 Strasbourg, France} 
\author{H. Lumia}
    \affiliation{NTT Research, Inc. Physics \& Informatics Laboratories, 940 Stewart Dr, Sunnyvale, CA 94085}%
    \affiliation{ESPCI Paris-PSL, 10 rue Vauquelin, 75231 Paris Cedex 05, France}     
\author{E. Ng}
    \affiliation{NTT Research, Inc. Physics \& Informatics Laboratories, 940 Stewart Dr, Sunnyvale, CA 94085}%
    \email{edwin.ng@ntt-research.com}
\author{T. Chervy}
    \affiliation{NTT Research, Inc. Physics \& Informatics Laboratories, 940 Stewart Dr, Sunnyvale, CA 94085}%
    \email{thibault.chervy@ntt-research.com}
\date{\today}

\begin{abstract}
We propose a time-multiplexed photonic network architecture based on coupled ring resonators, capable of accurately emulating specific Hamiltonian dynamics. We show that, in the Suzuki–Trotter limit, the resulting stroboscopic evolution reproduces the characteristic dynamics of the bosonized Hopfield model. Furthermore, by incorporating a nonlinear element within the main resonator loop, we outline a scalable route toward optical simulation of both mean-field and quantum nonlinear dynamics associated with the Tavis–Cummings model. Our results establish time-multiplexed resonator networks as a versatile photonic framework for simulating interacting light–matter Hamiltonians and collective many-body phenomena. 
\end{abstract}
\maketitle


\section{Introduction}

Optical analogue simulators have recently emerged as versatile platforms to study complex many-body systems~\cite{aspuru2012photonic, inagaki2016large, altman2021quantum, madsen2022quantum, Clark_Schine_Baum_Jia_Simon_2020, BenLev2025, calvanese2021all, berloff2017realizing}. On one hand, they provide precise control over microscopic degrees of freedom such as position, energy, momentum, and polarization of photons, allowing for precise initialization of the optical states. On the other hand, quantum optical toolbox enables direct measurement of correlation functions and photon statistics both in time and space, which are crucial in the identification of complex many-body phases.

Synthetic dimension networks offer a promising path toward large scale simulations of such systems. The key idea of such networks is to emulate a lattice in a so-called synthetic space that has larger scale and/or higher dimensionality than the physical device~\cite{Yuan_Lin_Xiao_Fan_2018_rev, Ozawa_Price_2019}. To this end, a set of physical modes is required, along with a mean of coupling them, to implement the structure of the lattice. In optics, different approaches have been explored to create synthetic lattices, such as frequency-multiplexing in multimode resonators \cite{roslund2014wavelength, lu2023frequency}, angular-momentum-multiplexing in ring cavities \cite{Ozawa_Price_Goldman_Zilberberg_Carusotto_2016}, or time-multiplexing in optical fiber and photonics networks~\cite{marandi2014network, chalabi2019synthetic}. 

In a time-multiplexed architecture, a single optical transverse mode is used to generate an arbitrary number of sites in the synthetic dimension. This particular mean of creating a synthetic network has proven successful to implement quantum simulators~\cite{Yuan_Xiao_Lin_Fan_2018_art, takeda2017universal}. Most notably, such networks feature no site-to-site variation and enable arbitrary connectivity~\cite{Marandi_2014, leefmans2021topological, bartlett2024programmable}. In addition, they offer direct access to the individual temporal sites, \textit{e.g.} through dynamical phase-shifters and time-resolved homodyne measurements, thereby opening propitious possibilities regarding many-body system simulation. However, experimental realizations of the coupling between different time-multiplexed sites have so far been limited to dissipative couplings using beamsplitters \cite{leefmans2021topological}, where vacuum noise addition and energy loss are incurred at every coupling event. As a result, the physics emulated so far has been restricted to mean field dynamics, where the loss associated with the couplers are off-set by added gain in the main cavity loop. While this approach captures very well mean field features, and allows the exploration of various non-Hermitian models, the extension of time-multiplexed networks to the quantum regime will require new architectures where losses and noise addition can be separated from site-to-site couplings.

In this paper we propose such an architecture, implementing a time-multiplexed network (TMN) analogue of the widely used bosonized Hopfield-like model~\cite{Ciuti_Bastard_Carusotto_2005, Todorov_Sirtori_2012, De_Liberato_Keeling_2016, George_Chervy_Shalabney_Devaux_Hiura_Genet_Ebbesen_2016, Herrera_Spano_2016,  Zeb_Kirton_Keeling_2018}. Our proposed scheme is based on two coupled ring-resonators, where the evanescent coupling between the two cavities populates time-encoded lattice sites at every round-trip. In the Trotterized limit, where coupling rates are smaller than the inverse cavity round-trip time, this platform successfully reproduces hallmarks of polariton physics, both in the conservative and driven-dissipative regimes. In particular, our simulator accurately reproduces the avoided crossing behavior of collective polaritonic states in the strong coupling regime, the $\sqrt{N}$-scaling of the normal mode splitting, as well as the brightening of dark states in the presence of a controlled on-site energy disorder. We demonstrate that the addition of a weakly nonlinear element in the main loop resonator results in a polaritonic bistability behavior, and we discuss possible extensions towards the simulation of quantum nonlinear dynamics. The operating regime of this resource-efficient simulator is compatible with state-of-the-art silicon photonics and telecom fiber technologies, and can be transposed to superconducting systems using microwave cavities and a single Josephon junction qubit as the source of strong nonlinearity. The proposed platform holds promises for the study of large-scale polaritonic systems with precise and dynamical control over the system's parameter down to the single site level, with implication for the study of polaritonic transport in disordered materials, quench dynamics, and quantum many-body physics.

\subsection*{Outline}

In Section~\ref{Sec:II_network_to_simulator} we detail the architecture and operating principle of the TMN and expose a theoretical framework that enables us to highlight the bridge between our synthetic network and Hopfield-like bosonic Hamiltonians. We then illustrate the performance of this platform with a numerical implementation of the aforementioned theory, before exploring practical realization of the TMN using well established optical elements in Section~\ref{sec:IV_real_device}. We then discuss possible strategies to implement nonlinearities in the system up to the fermionized limit of the TMN in Section~\ref{Sec:V_non_linear_simulator}.We draw our conclusions in Sec~\ref{Sec:Conclusion}.


\section{From A Resonator Network to a Quantum Simulator}
\label{Sec:II_network_to_simulator}

\subsection{\label{Sec:II_A}Building a Time-multiplexed Photonic Resonator Network (TMN)}

\begin{figure*}[htbp]
    \includegraphics{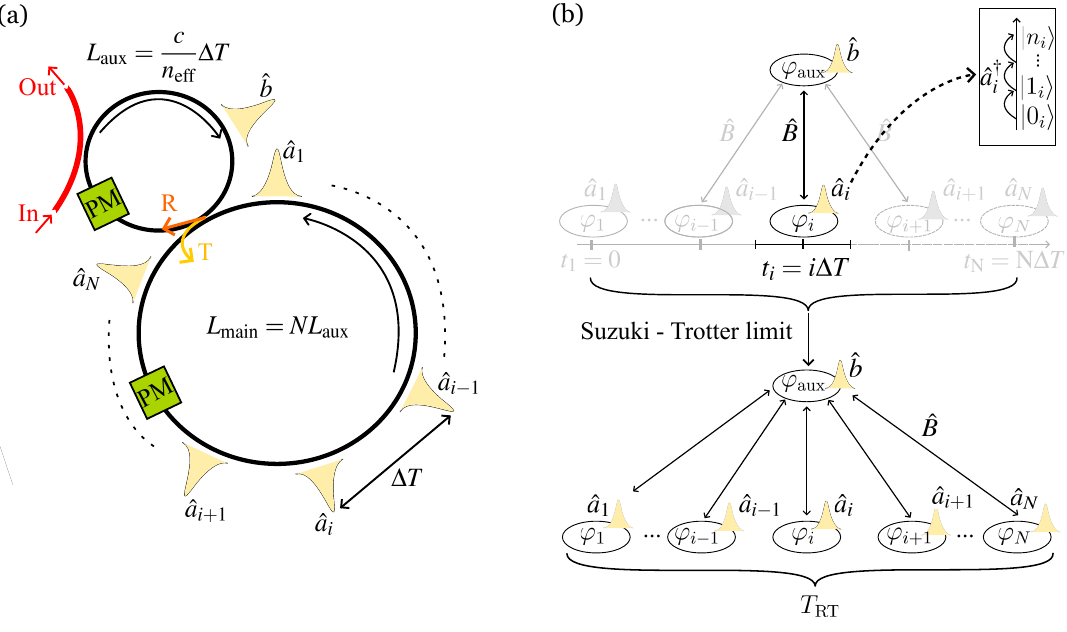}
    \caption{ Time-multiplexed photonic resonator network. (a) Sketch of the proposed architecture. An injection fiber loads a coherent state into the auxiliary cavity. At the evanescent coupling, a proportion $T^2$ of the light intensity is transmitted to the main cavity and the rest ($R^2$) continues to travel inside the auxiliary cavity. The length of the main cavity is N times larger than the one of the auxiliary cavity, defining N temporal sites inside the main cavity, achieving time-multiplexing of the main cavity transverse mode. (b) Representation of the temporal sites in the synthetic dimension. Each site $\hat a_i$ exchanges energy with the auxiliary cavity at a given time $t_i$, via a beam-splitter like operation. Each and every sites can store many photons, whose phase can be individually tuned using the intra-cavity phase modulators. One complete loop inside the main cavity is achieved in a time $T_\mathrm{RT} = N \times \Delta T$. In the Trotterized limit, every operation can be considered to happen simultaneously during the coarse-grained integration time $T_\mathrm{RT}$.}
    \label{figure1}
\end{figure*}

\bigbreak
Our proposed architecture is depicted in Fig.~\ref{figure1}(a). It consists of two evanescently coupled ring resonators of different sizes, each supporting a single transverse cavity mode. The longer (shorter) resonator is referred to as the "main cavity" ("auxiliary cavity"). Superpositions of longitudinal modes in each resonator allow the formation of optical pulses, which can couple back and forth between the two cavities at every round-trip. In the synchronous pumping regime, where the main cavity is exactly N times longer than the auxiliary cavity, a single pulse initialized in the auxiliary cavity generates a well-defined pulse train in the main cavity (Fig.~\ref{figure1}(a)). The time interval $\Delta T$ of the pulse train is imposed by the round-trip time of the auxiliary cavity. In the synthetic dimension, each pulse of the main cavity corresponds to a single lattice site, periodically exchanging energy with the auxiliary resonator. The phase of each pulse in the main and auxiliary cavities can be controlled by phase-modulators (PM). 

In the absence of nonlinearity, each temporal site is described by a bosonic operator acting on an infinite ladder of Fock states (Fig.~\ref{figure1}(b)). We define $\{\hat a_i^\dagger\}_{i \in [\![ 1, N ]\!] }$ and $\hat b^\dagger$ as the creation operators for temporal sites in the main and auxiliary cavities, respectively. For simplicity, we use $\hbar = 1$ in the whole text. 

The coupling between the main cavity sites $\hat a_i$ and the auxiliary site $\hat b$, can be described by a beam-splitter operation with a mixing angle $\theta_\mathrm{disc}$, 
\begin{equation}
    \label{eq:BS}
    \hat B_i = e^{\theta_\mathrm{disc} (\hat a_i \hat b^\dagger - \hat a_i^\dagger \hat b)}.
\end{equation}
The angle $\theta_\mathrm{disc}$ quantifies how much energy is exchanged at each round-trip between the auxiliary cavity and the main cavity, corresponding to the round-trip integrated coupling operation as the pulses traverse the evanescent coupling region~\cite{Haroche2006exploring}.
The effect of intra-cavity phase shifters are described by :
\begin{eqnarray}
    \hat P_\im &= e^{-i \varphi_{i} \hat a_i^\dagger a_i} \nonumber \\
    \hat P_\mathrm{aux} &= e^{-\im \varphi_\mathrm{aux} \hat b^\dagger b}.
\end{eqnarray}

Similarly, the phase-shifts $\varphi_i$ are roundtrip integrated quantities, related to the time spent by the pulses in the phase shifters. Collecting these unitaries, we build the round-trip operator of the TMN as
\begin{equation}
    \label{eq:sequential_U_1}
    \hat{\mathcal U}_\mathrm{rt} = \Pi_{\im=1}^N \left( \hat P_\im \hat B_\im \hat P_\mathrm{aux}\right).
\end{equation}

We now turn to the description of drive and dissipation in the TMN. In our architecture, a loading fiber, depicted in red in Fig.~\ref{figure1}, is used to inject light in the auxiliary cavity at every roundtrip $l$ with a controlled phase and amplitude. Such a driving scheme is modeled using 
\begin{equation}
\label{eq:udrive_final_main}
    \hat U_\mathrm{drive}(l) = \exp\left\{ -\im\left( F_\mathrm{disc}\mathrm{e}^{\im \ell \varphi_\mathrm{pump}}\hat b + F_\mathrm{disc}\mathrm{e}^{-\im \ell \varphi_\mathrm{pump}}\hat b^\dagger\right) \right\},
\end{equation}
where $F_\mathrm{disc}$ is the round-trip integrated driving amplitude, and $\ell \varphi_\mathrm{pump}$ is a round-trip integrated phase, emulating a continuous driving in the synthetic space (see Supplementary Material Sec.I~\cite{SuppMat}). In our numerical simulations, we considered a strongly under-coupled auxiliary cavity, driven once before each coupling event. A more refined approach would consist in using a controlled injection inside the auxiliary cavity via a non-linear coupling element~\cite{heuck2020controlled, Bustard_Bonsma-Fisher_Hnatovsky_Grobnic_Mihailov_England_Sussman_2022, Karni_Vaswani_Chervy_2025}.

Rigorously, our system is subjected to continuous propagation losses. In the regime where these losses are small as compared to the other round-trip coupling rates, we can collect all the round-trip losses for each site $i$ into a single round-trip integrated dissipator, $\hat{{L}}_i = \sqrt{\gamma_\mathrm{rt}} \hat a_i$ where $\gamma_\mathrm{rt} = \gamma T_\mathrm{RT}$, and $\gamma$ accounts for intrinsic losses of the cavity, as well as potential photon loss due to the various optical elements inside the loop.

In Sec.~\ref{Sec:V_non_linear_simulator} we explore the effect of a $\chi^{(3)}$ non-linearity on the dynamics of our time-multiplexed network. Several way of implementing this non-linearity are possible, as described further in the aforementioned section. In order to add the non-linear term in our model, we need to supplement the round-trip operator Eq.~\eqref{eq:sequential_U_1} with 
\begin{equation}
    \label{eq:disc_nl_hamiltonian}
    \hat U_\mathrm{Kerr}^i = e^{U_\mathrm{Kerr} \hat n_i (\hat n_i - 1)}.
\end{equation}
The effect of applying this unitary to an optical mode has been studied in depth in \cite{Kitagawa_Yamamoto_1986}. It can be described as a self-phase modulation effect, where the phase of the traveling mode is modulated by its own intensity. As seen in the exponent of Eq.\eqref{eq:disc_nl_hamiltonian}, a state populated by more than one photon will experience an energy shift of order $U_\mathrm{Kerr}$. Note that $U_\mathrm{Kerr}$ is the round-trip integrated nonlinear interaction, related to the time spent by the pulses inside the nonlinear medium~\cite{Kitagawa_Yamamoto_1986}.

At this stage, Eq.~\eqref{eq:sequential_U_1} realizes \textit{sequential} energy exchanges between the auxiliary resonator and the main cavity sites. We now expose how, in the Susuki-Trotter limit, this model maps to the continuous evolution of a Hopfield-like Hamiltonian.

\bigbreak

\subsection{Mapping to Hopfield-like bosonic models}

\label{sec:mapping}
To simplify the discussion, we focus in this section on the conservative terms of the TMN evolution operator (Eq.~\eqref{eq:sequential_U_1}). The non-conservative  terms transforms analogously \cite{Campaioli_Cole_Hapuarachchi_2024}. 

The key idea of our TMN emulation of a continuous Hamiltonian is rooted in the Susuki-Trotter (ST) approximation of the round-trip evolution operator,
\begin{equation}
\label{eq:ST_2}
    \hat{\mathcal U}_\mathrm{RT} =\Pi_{i=1}^N \left\{\exp{\left(\delta A_i\right)} \right\} = \exp{\left(\delta \sum_{i=1}^N A_i\right)} + \mathcal{O}(\delta^2).
\end{equation}
In the ST approximation, the round-trip evolution is thus given by a continuous-time Hamiltonian
\begin{equation}
    \label{eq:H_ST}
    \mathcal{H}_{ST} = \frac{\varphi_\mathrm{aux}}{T_\mathrm{RT}} \hat b^\dagger \hat b + \sum_{i=1}^N \frac{\varphi_i}{T_\mathrm{RT}} \hat a_i^\dagger \hat a_i^{\phantom{\dagger}} + \frac{\theta_\mathrm{disc}}{T_\mathrm{RT}}\sum_{i=1}^N (\hat a_i^\dagger \hat b + b^\dagger \hat a_i^{\phantom{\dagger}}),
\end{equation}
acting simultaneously on all the degrees of freedom of the system, with quantities whose magnitudes are integrated over a round-trip. The round-trip time $T_\mathrm{RT}$ can be seen as a coarse-grained quantity, or integration time, during which all applied operations are considered to be simultaneous.
Intuitively, the ST limit corresponds to the high finesse limit of the auxiliary cavity, where the probability of energy exchange at each round-trip is small~\cite{PhysRevResearch.4.013009}. In this setting, the time-ordering of the pulses reaching the coupling region is irrelevant, and all pulses experience the same auxiliary cavity mode population. We further analyze in Section \ref{Sec:ST_limit} and Supplemental Material~\cite{SuppMat} the impact of the coupling strength, phase shifts, driving, and dissipation on the validity of the ST approximation. 

Comparing the ST Hamiltonian Eq~\eqref{eq:H_ST} with the familiar bosonic Hopfield model
\begin{equation}
    \label{eq:H_B}
    \mathcal H_{B} = \omega_c \tilde b^\dagger \tilde b + \sum_{i=1}^N \omega_{i} \tilde a^\dagger_i \tilde a_i^{\phantom{\dagger}} + g\sum_{i=1}^N  (\tilde a_i^\dagger \tilde b + \tilde b^\dagger \tilde a_i^{\phantom{\dagger}}), 
\end{equation}
where $\omega_c$ and $\omega_i$ are the cavity and (bosonized) material modes energies and $\Omega_R = g\sqrt{N}$ is the collective light-matter coupling strength,
we establish the correspondence between the simulation parameters and the Hopfield-like quantities in TABLE~\ref{table1}.

\begin{table}[h]
\begin{tabular}{ | c | c |}
\hline
Simulation & Hopfield-like \\
\hline
Phase shift & Energy \\
$\varphi_i/T_\mathrm{RT}$ & $\omega_i $ \\
\hline
Beamsplitter angle & Coupling strength \\
$\theta_\mathrm{disc}/T_\mathrm{RT}$ & $g$ \\
\hline
Simulation time & Emulated time \\
$n \times T_\mathrm{RT}$ & $ t $  \\
\hline
\end{tabular}
\caption{\label{table1}%
Correspondence between the simulation parameters and the emulated model parameters.
}
\end{table}

Our TMN simulator thus emulates, in the ST limit, the continuous dynamics of the Hamiltonian Eq.~\eqref{eq:H_ST}. It is a fully bosonic Hamiltonian that represents one bosonic mode at energy $\varphi_\mathrm{aux}/T_\mathrm{RT}$ interacting continuously with N other bosonic modes with energies $\varphi_i/T_\mathrm{RT}$. Such a bosonized picture is widely used in the field of light-matter interaction and is at the roots of the historical diagonalization of the strong coupling Hamiltonian by J. J. Hopfield \cite{Hopfield_1958} and more recently used to describe organic molecules in cavity \cite{De_Liberato_Keeling_2016, George_Chervy_Shalabney_Devaux_Hiura_Genet_Ebbesen_2016, Herrera_Spano_2016, del_Pino_Feist_Garcia-Vidal_2015, Strashko_Keeling_2016, Zeb_Kirton_Keeling_2018} as well as excitons in semiconductors \cite{Ciuti_Bastard_Carusotto_2005, Todorov_Sirtori_2012}.

\bigbreak
\section{\label{Sec:simulation_results} Simulation results}

In the following section, we detail the procedure of the numerical simulation and establish the regime of validity of the ST approximation, in which our system is expected to correctly reproduce well known features of Hopfield-like models. The numerical implementation is done in Julia, using the package \textit{QuantumOptics.jl} \cite{kramer2018quantumoptics}.

\subsection{\label{Sec:num_imp}
Numerical procedure}

In this work, two different kinds of numerical simulations are presented. The first is the conservative limit of the TMN, whose results are presented in Sec.~\ref{Sec:ST_limit}, based on the roundtrip evolution Eq.~\eqref{eq:sequential_U_1}. It is used to provide a quantitative analysis of the ST limit, in a regime where losses are ignored, and allows for a simple definition of the simulation fidelity, as explained in Sec.~\ref{Sec:ST_limit}, as well as a study of the scaling of the Rabi frequency with the number of temporal sites in Sec.~\ref{Sec:Rabi_results}. The second regime of interest is the driven-dissipative regime of the TMN, exposed in Sec.~\ref{Sec:Rabi_results}, based on a complete round-trip evolution operator, including driving through Eq.\eqref{eq:udrive_final_main} and propagation losses.

In the driven-dissipative regime, we start from a vacuum density-matrix defined as $\rho_0 = |\psi_0 \rangle \langle \psi_0 |$ where $|\psi_0 \rangle = |0_\mathrm{aux} \rangle \otimes_{i=1}^N [0_i \rangle$ corresponds to all sites empty of photons. 
We then set every discrete quantities, $\{\varphi_\mathrm{aux}, \varphi_i, \theta_\mathrm{disc}, F_\mathrm{disc}, \varphi_\mathrm{pump}, \gamma_\mathrm{RT} \}$ which are next used to create all the Hamiltonians and dissipators needed for the round-trip evolution. For simplicity, we set our reference unit of time as $T_\mathrm{RT} = 1$. The system is now fully defined and initialized. The dynamics of the TMN is numerically simulated by sequential evolution of the initial density matrix with all the previously defined Hamiltonians and dissipators, until the desired number of roundtrips is reached. The reference state is computed from the same initial density matrix by applying the ST Hamiltonian Eq.~\eqref{eq:H_ST} supplemented with a continuous driving term and dissipation, using a built-in Lindblad master equation solver.

In the conservative limit, the same procedure is applied, but the initial state is chosen to be $|\psi_0\rangle = |\alpha_\mathrm{aux} \rangle \otimes_{i=1}^N [0_i \rangle$, where |$\alpha_\mathrm{aux} \rangle$ is a coherent state injected inside the auxiliary cavity site with $\sqrt{\alpha}$ mean photon number. Both the reference and the TMN are evolved for a suitable time, using only the Hamiltonian parts, devoided of the driving terms.
In both case we extract the population and desired expectations values on the temporal sites after each roundtrip.

\subsection{Quantitative analysis of the ST limit}
\label{Sec:ST_limit}
\begin{figure*}[htb!]
    \centering
    \includegraphics{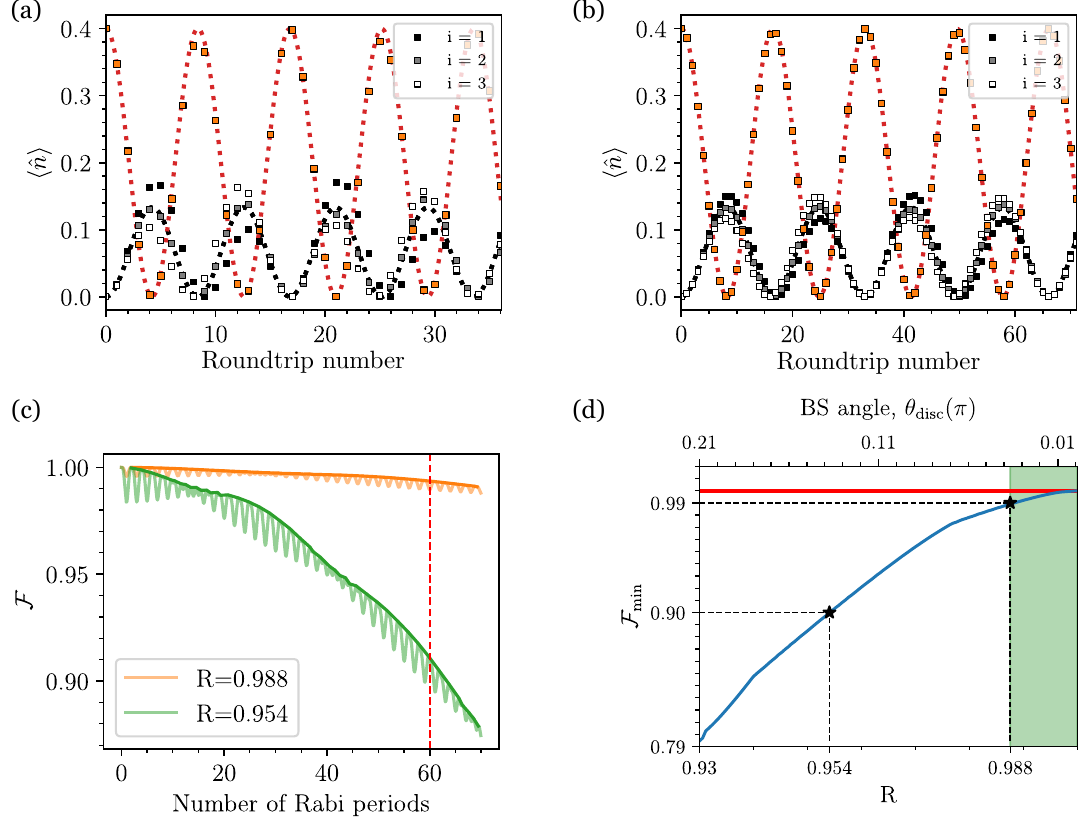}
    \caption{Dynamics and fidelity of the TMN simulator. In all simulation N=3 and the dynamics are initialized with $|\alpha = \sqrt{0.4} \rangle$ in the auxiliary cavity. (a)~\&~(b) Time evolution of the mean population for the TMN (squares) and continuous (dotted line) model. Red (orange) color labels the auxiliary cavity site and shades of grey the main cavity sites. Reflectivity is $R = 0.954$ for (a) and $R = 0.988$ for (b). (c) Time evolution of the Fidelity as a function of the number of Rabi periods elapsed. The two curve corresponds to long-time dynamics of Reflectivity (a)~\&~(b). The red dashed line indicate the time cut-off used in (d). (d) Evolution of the global minimum of the fidelity, after 60 Rabi oscillations, as a function of Reflectivity (or BS angle). Red horizontal line shows the maximal $\mathcal{F} = 1$ value of the fidelity.}
    \label{figure2}
\end{figure*}

As explained in the preceding section, the analysis of the ST limit is done in the conservative framework. This choice is motivated by both computational and physical reasons. On one hand, the metric we use to assess the precision of the simulation is most easily computed using a conservative system (see the discussion about the fidelity Eq.~\eqref{eq:fidelity} below). On the other hand, as we will be focusing on the strong coupling regime of the bosonic Hopfield Hamiltonian, we restrict ourselves to $\theta_\mathrm{disc} > \gamma_\mathrm{RT}$ (an in depth discussion about losses is provided in Sec.~\ref{sec:IV_real_device}), and thus one can disregard losses in this section.
However, rigorously, the ST limit should also be justified in terms of the other discrete quantities, namely phase shifts and driving. The interesting space of parameters for this study is restricted to near-resonant interactions and weak external driving ($\varphi_\mathrm{aux}-\varphi_i \sim \sqrt{N}\theta_\mathrm{disc}$, $\varphi_\mathrm{pump} - \varphi_i \sim \sqrt{N}\theta_\mathrm{disc}$  and $\theta_\mathrm{disc} > F_\mathrm{disc}$). Keeping $\sqrt{N}\theta_\mathrm{disc}/T_\mathrm{RT}$ as the largest energy scale of our system, we only need to inspect the ST limit in terms of this parameter.  We further analyze in Supplemental Material~\cite{SuppMat} the impact of the other simulation parameters on the fidelity of the emulation, including the number of pulses simultaneously present in the system.

Let us now discuss the high-finesse limit for the auxiliary cavity. We fix the number of temporal sites to $N=3$ to reduce the computation runtime. We also set the same energy for every mode and thus $\varphi_\mathrm{aux} = \varphi_i = \varphi$. 
The fidelity of the simulator is then defined, following Ref.\cite{Casanova_2012}, as
\begin{equation}
\label{eq:fidelity}
    \mathcal F_0 = |\langle \psi_\mathrm{cont}(t_0)|\psi_\mathrm{disc}(l_0)\rangle|^2~,
\end{equation}
where $t_0 = l_0 \times T_\mathrm{RT}$ and $ |\psi_{i}(t)\rangle$ is a many-body state of the system as described in \ref{Sec:num_imp}, taken at time $t$.

We show in Fig.~\ref{figure2}~(a,b) the evolution of the average photon number on each site of the main and auxiliary cavities, as a function of the number of resonator round-trips, for different values of the beam splitter reflection coefficient, together with the continuous evolution of the ST limit Hamiltonian. Direct comparison of these two graphs shows that the larger coupling strength of Fig.~\ref{figure2}~(a) ($R = 0.954$) results in larger deviations from the ST limit Hamiltonian than the weaker coupling case of Fig.~\ref{figure2}~(b) ($R = 0.988$). Note that the discrete simulation results follow periodic rephasings to the continuous model every second Rabi periods (\textit{e.g.} around round-trip numbers 16-18 in panel (a) and 26-28 in panel (b)). As shown in Supplementary Material~\cite{SuppMat}, this effect only occurs at zero detuning, and originates from the sequential interaction of the main cavity pulses with the auxiliary cavity.

The evolution of the fidelity Eq.~\eqref{eq:fidelity} with simulation time, normalized by the Rabi period, is shown in Fig.~\ref{figure2}~(c) for two different values of the beamsplitter angle, $\theta_\mathrm{disc} \sim 0.14 \pi (R=0.954)$  and $\theta_\mathrm{disc} \sim 0.04 \pi$. The oscillatory dynamic of the fidelity is a result of two competing periodic behaviors: the above mentioned rephasing at twice the Rabi period, and a weaker oscillation at the Rabi frequency between maximum and minimum fillings of the main cavity sites. These competing behaviors are studied more in depth in the Supplementary Material. Besides these oscillatory features, the overall evolution of the fidelity shows a decaying trend, as expected from the accumulation of errors in the simulator. As shown in Fig.~S1 the leading contribution to these errors is due to an accumulated lag of the discrete TMN with regard to the continuous model, potentially amendable to simple error correcting schemes. 

Finally, we show in Fig.~\ref{figure2}~(d) the evolution of the minimal fidelity $\mathcal F_\textrm{min}$ of the simulator as a function of the beam splitter coupling strength. This quantity is defined as the minimal value taken by the fidelity over a simulation time of 60 Rabi periods. This value is chosen as a generally sufficient number of roundtrips to reach steady-state in a driven-dissipative regime. As indicated by the green region on this graph, our simulator exceeds $99\%$ minimum fidelity for beam splitter reflectivities above $R>0.988$.

\subsection{Accurate emulation of Hopfield-like model}
\label{Sec:Rabi_results}
\begin{figure*}
    \centering
    \includegraphics{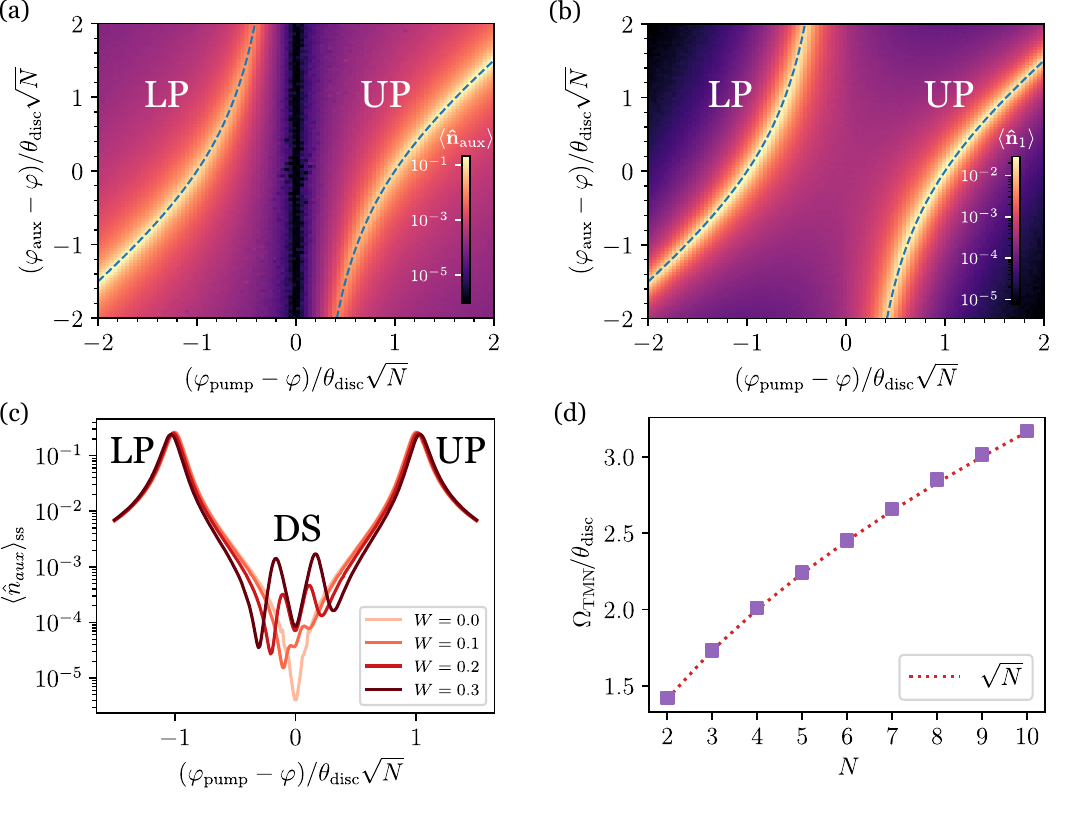}
    \caption{Numerical simulation of Hopfield-like model emulation. Here we take the reference of phases as $\varphi = 0$. For all simulation $\theta_\mathrm{disc} \sim 0.03 \pi~(R=0.99)$, $\theta_\mathrm{disc}/\gamma_\mathrm{RT} = 10$, $F_\mathrm{disc}/\gamma_\mathrm{disc} = 0.1$.(a) Steady-state population of the auxiliary cavity site for varying  main cavity/pump and main/auxilliary cavities detuning. The two bright branches corresponds to lower (LP) and upper (UP) polaritons. Dashed line corresponds to Eq.~\eqref{eq:polaritons_dispersion}. (b) Same as (a) for the first main cavity sites.  (c) Steady state auxiliary cavity population for several values of main cavity pulses diagonal disorder. Main cavity pulses energies are uniformly distributed in $[\varphi - W, \varphi + W]$. (d) Evolution of the Rabi-splitting of the TMN ($\Omega_\mathrm{TMN}$) with the number of pulses fitting inside the main cavity. The red dashed line represented the theoretical curve $\omega_\sim/\theta_\mathrm{disc} = \sqrt{N}$ curve.} 
    \label{figure3}
\end{figure*}

All the subsequent results are obtained in the driven-dissipative regime. The Hamiltonian Eq.~\eqref{eq:H_ST} is easily diagonalized in the Fourier basis and in the 1-excitation subspace. Its eigen spectrum is composed of two polaritonic states \cite{Haroche2006exploring}, or bright states, which can be interpreted as hybrid states where a quantum of energy is shared between all the main cavity sites and the auxiliary cavity. The remaining $N-1$ dark states involves collective excitations of the main cavity sites only. In our formalism, the energies of the hybrid states are written (omitting the $T_\mathrm{RT} = 1$ factor and assuming $\varphi_i = \varphi~\forall~i~\in~\llbracket 1,N \rrbracket$):  

\begin{eqnarray}
\label{eq:polaritons_dispersion}
    E_\mathrm{UP/LP} &=& \frac{\varphi + \varphi_\mathrm{aux}}{2} \pm \sqrt{\left(\theta_\mathrm{disc}\sqrt{N}\right)^2 + \left(\frac{\varphi_\mathrm{aux} - \varphi}{2}\right)^2} \nonumber \\
\end{eqnarray}
We show in Fig.~\ref{figure3}~(a) the steady-state population of the auxiliary cavity for varying detuning and drive frequencies, together with the modeled polariton dispersion Eq.~\eqref{eq:polaritons_dispersion}. In the Hopfield language, this measurement maps the photonic content of the polaritonic states, analogous to a spectroscopic cavity measurement. We present in Fig.~\ref{figure3}~(b) the corresponding dispersion diagram for the average steady-state population of the main cavity sites, which features the spectral signature of both bright and dark states. In the polaritonic language, this measurement maps the matter-like excitation content of the hybrid states.

As mentioned above, dynamical phase control of the main cavity sites allow to emulate energetic disorder in the material system. We show in Fig.~\ref{figure3}~(c) the average stead-state population of the auxiliary cavity site for different strengths of a uniformly distributed diagonal disorder on the main cavity site energies. As expected, when $W=0$ we recover the spectroscopic signature of the two bright polaritonic states. Upon increasing the disorder strength, we observe the brightening of the dark states, emerging near the uncoupled site energies \cite{Gera_Sebastian_2022a, Gera_Sebastian_2022b, Sun_Dou_Gelin_Zhao_2022}.

Another hallmark of Hopfield-like models is the collective enhancement of the Rabi splitting, reflected in the $\sqrt{N}$ scaling of the splitting energy with the number of coupled emitters. We extract the Rabi-splitting from the Fourier transform of a conservative evolution over 10 Rabi periods. The use of the conservative model for this simulation is motivated by the exponential scaling of the matrix sizes with the number of sites in an open-system framework. Repeating this procedure for different number of pulses $N$, we demonstrate in Fig.~\ref{figure3}~(d) a close agreement with the expected collective strong coupling behavior. The above results establish our TMN architecture as an accurate simulator of Hopfield-like models in the high-finesse (or Suzuki-Trotter) limit.

\section{\label{sec:IV_real_device}Device fabrication perspectives}

\begin{figure}
    \includegraphics[width = \columnwidth]{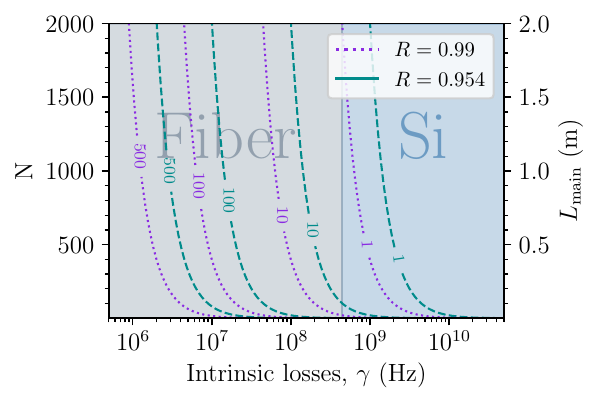}
    \caption{ Cooperativity diagram of a device with a fixed auxiliary cavity length of $L_\mathrm{aux} = 1\,$mm and for various $\theta_\mathrm{disc}$ angles. Here $n=1.5$ . All the part to the left of the cooperativity = 1 line corresponds to the strong coupling domain. The silver part correspond to the optical fiber domain and the light blue part to the Si integrated photonics domain. The left edge of the Si domain indicates the limit of state-of-the-art cavities.}
    \label{figure4}
\end{figure}

We now turn to the question of the maximum system size that can be emulated by our TMN approach, under realistic device constraints. To this end, let us first reemphasize that, in this architecture, the auxiliary cavity round-trip time sets the temporal spacing between pulses in the main cavity Fig.~\ref{figure1}. In order to increase the system size, a simple approach thus consists in shrinking the auxiliary cavity length, and increasing the repetition rate of the pump accordingly, thus populating more temporal sites in the main cavity loop. This approach is ultimately limited by the bending losses of the auxiliary resonator. Considering typical photonic integrated circuit constraints, we set the lower limit of the auxiliary cavity length to ca. $1\,$mm, corresponding to a repetition rate of ca. 200\,GHz for a thin film Lithium Niobate device \cite{he2019self}. 

Given a length of auxiliary cavity, the simulated system size can further be tuned by increasing the length of the main cavity loop. This approach is in turn limited by the increasing roundtrip propagation losses $\gamma_{\textrm{RT}}$. The cooperativity of the light-matter coupling $\mathcal{C} = \Omega_\mathrm{R}/\gamma$ is then expressed as:
\begin{equation}
    \mathcal{C} = \frac{\theta_\mathrm{disc}\sqrt{N}}{\gamma_\mathrm{RT}} = \frac{\theta_\mathrm{disc}c}{\gamma \sqrt{N}L_\mathrm{aux}n},
\end{equation}
where $c$ is the speed of light, $n$ the refractive index, and we used $L_\mathrm{main} = N \times L_\mathrm{aux}$ and $T_\mathrm{RT} = nL_\mathrm{main}/c$. The surprising scaling of $\mathcal{C}$ with $N^{-1/2}$ arises from the competition between collective coupling and increased propagation losses for larger system sizes, accommodating more temporal sites. 

We show in Fig.~\ref{figure4} the evolution of the cooperativity for various beamsplitter mixing angles, as a function of the number of temporal sites and intrinsic loss rate, for an auxiliary cavity length of 1\,mm. The colored regions in this diagram correspond to the achievable intrinsic loss rates reported for integrated silicon photonics (light blue), corresponding to Q-factors around $10^6$ ~\cite{Shankar_Bulu_Loncar_2013, Luke_Okawachi_Lamont_Gaeta_Lipson_2015, Miller_Yu_Ji_Griffith_Cardenas_Gaeta_Lipson_2017,  Armand_Perestjuk_Della_Torre_Sinobad_Mitchell_Boes_Hartmann_Fedeli_Reboud_Brianceau_et_al._2023} and losses of c.a. 0.2 dB/cm, and for telecom optical fibers (gray) with propagation losses under 0.2 dB/km~\cite{Sakr_Chen_Jasion_Bradley_Hayes_Mulvad_Davidson_Numkam_Fokoua_Poletti_2020}. Thus, both platforms can accurately emulate high-cooperativity, large system-size Hopfield-like models under realistic device constraints.

\section{\label{Sec:V_non_linear_simulator}Towards Non-linear Quantum Simulator}

\bigbreak

\begin{figure*}
    \includegraphics{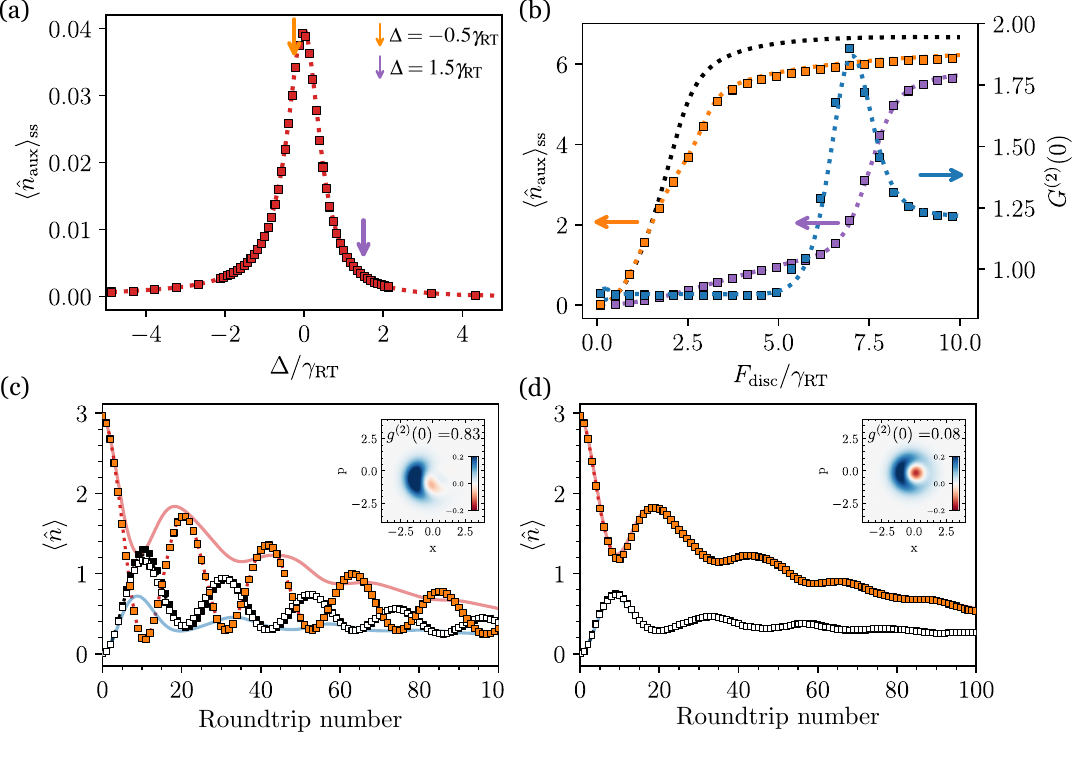}
    \caption{Non linearity and qubit limit for the TMN simulator. Dotted Lines represents the ST Hamiltonian Eq.~\eqref{eq:H_ST}, continuous faded lines the ST Hamiltonian in the qubit limit and square symbols the TMN. Red (orange) colour labels the auxiliary cavity population and shades of grey the main cavity one.  Simulation are done at $\theta_\mathrm{disc} \sim 0.0.3\pi~$($R = 0.99$) and $\theta_\mathrm{disc}/\gamma_{RT} = 5$. When not specified, $F_\mathrm{disc}/\gamma_{disc} = 0.1$. (a) Spectroscopy of the lower polariton of the system for N = 1. Colored arrows corresponds to corresponding curves on (b). Detuning are expressed in units of $\gamma_\mathrm{RT}$. (b) Auxiliary cavity population at steady-state for increasing pump amplitude in the optical limiter (orange) and bistable (purple) cases. Black dotted line represents the same detuning as the optical limiter case, with no non-linearity. Saturation behaviors at high pumping amplitude is dominated by truncation effects, as the Fock spaces are truncated to $|n_\mathrm{max}\rangle = |12\rangle$. (c) Dynamic  of the quantum simulator and the ST Hamiltonian until steady-state for $U_\mathrm{Kerr} = \theta_\mathrm{disc}$. Insert are contour plots of the Wigner function for the main cavity site taken at maximum filling (around roundtrip 10). (d) Same as (c) for $U_\mathrm{Kerr} = 10~\theta_\mathrm{disc}$ \textit{ie} in the fermionized (or qubit) limit. (c) \& (d) are done with N = 2.}
    \label{figure5}
\end{figure*}

The results shown above demonstrate that our TMN platform successfully emulates Hopfield-like Hamiltonians in the dilute (bosonized) limit. In this section, we propose a path towards emulating quantum effects, which are manifest in the presence of strong on-site nonlinearities, for which the bosonic approximation of light-matter Hamiltonians fails. As an intermediate step towards this frontier, we also demonstrate that meanfield nonlinear effects can be emulated in the weakly interacting regime. 

In order to implement photon-photon interactions on each of the temporal sites, a non-linear element can be added inside the main cavity loop. 
We focus here on self-phase modulation interactions, based on a $\chi^{(3)}$ nonlinearity, in which each pulse undergoes a phase-shift with a magnitude that depends on its instantaneous population \cite{Kitagawa_Yamamoto_1986}). As explained in \ref{Sec:II_A}, numerical simulations of the Kerr effect are performed by supplementing the round-trip unitary with the non-linear evolution operator Eq.~\eqref{eq:disc_nl_hamiltonian}. 

In the regime of weak nonlinearities ($U_\mathrm{Kerr} < \gamma_\mathrm{RT} < \theta_\mathrm{disc}$), 
the TMN is expected to capture driven-dissipative dynamics, similar to those observed at the mean-field regime in \textit{e.g.} quantum well polaritonic systems \cite{Carusotto_Ciuti_2013}. We show in Fig.~\ref{figure5}~(b) the evolution of the steady-state population of the auxiliary cavity, for different detunings $\Delta = \varphi_\mathrm{pump} - E_\mathrm{LP}$ between the pump and the lower polariton mode, indicated by matching color arrows on Fig.~\ref{figure5}~(a). Here, we set the number of temporal sites to N=1 in order to maximize the size of the numerical Fock space ($|n_\mathrm{max}\rangle = |12\rangle$), while keeping reasonable simulation run time. As expected from meanfield calculations~\cite{rodriguez2017probing, fink2018signatures}, a red-detuned drive of the lower polariton branch (orange curve) results in an optical limiter effect, with a steady-state population growing sublinearly with the pumping strength. At positive detunings $\Delta > \sqrt{3}/2 \gamma_\mathrm{RT}$, (purple curve) the system displays the well-known bistable behavior, accompanied by a strong bunching peak in the photon number statistics of the auxiliary cavity mode at the critical point of the dissipative phase transition (blue curve)~\cite{fink2018signatures}. We note however that the numerical Fock space truncation impacts the behavior of these simulations at high average photon numbers. This is evidenced in Fig.~\ref{figure5}~(b), black dotted line, which represents the evolution of the steady-state population at $\Delta/\gamma_\mathrm{RT} = -0.5$ in the absence of non-linearities. The saturation behavior of this curve above $\langle \hat n_\mathrm{aux} \rangle_\mathrm{ss} \sim 5$ is a result of the truncation of the numerical Fock space.

Augmenting our simulator with a weakly nonlinear material thus allows to emulate key hallmarks of the meanfield nonlinear polariton physics, enabling the study of driven-dissipative phase transitions with exquisite control over to the system parameters, such as disorder, dimensionality, individual site occupancy, and detuning, as discussed above.

As the strength of nonlinearity increases, quantum effects are expected to become manifest. We show in Fig.~\ref{figure5}~(c) the temporal evolution of the system (square symbols), following an initial injection of a coherent state with $\langle\hat{n}\rangle=3$ in the auxiliary cavity, coupled to two main cavity site experiencing a nonlinearity $U_\mathrm{Kerr} = \theta_\mathrm{disc} = 0.1\,\textrm{rad}$. While this time evolution closely follows the continuous model dynamics (dotted lines), it significantly deviates from the evolution of a continuous Hamiltonian in the qubit limit, corresponding to the regime of impenetrable bosons~\cite{Carusotto_Gerace_Tureci_De_Liberato_Ciuti_Imamoglu_2009, mabuchi2012qubit}, \textit{i.e.} with a Fock space truncated to $\{|0\rangle ,|1\rangle\}$ in the main cavity photon number basis (continuous curves). Nevertheless, computing the Wigner function of the system at the maximum filling of the main cavity site (roundtrip number 10) reveals the onset of negativity, as shown in inset. Such Wigner function negativity is a clear signature of non-classical behavior of the simulator. Computing the zero-delay second-order correlation function for this state of the cavity yields a value of $g^{(2)}(0) = 0.85$, further confirming the onset of quantum effects in this nonlinear regime.

Increasing the strength of the nonlinearity to $U_\mathrm{Kerr} = 10 \theta_\mathrm{disc}$ brings the system in a strongly anharmonic regime, as shown in Fig.~\ref{figure5}~(d). In this extreme regime, the presence of a single photon in the main cavity site blockades the injection a second photon, achieving $g^{(2)}(0) = 0.096$ and a clear negativity of the Wigner function (see inset of Fig.~\ref{figure5}~(c)). It should however be noted that the exact field dynamics in such strongly nonlinear regimes is far more complex than what our model captures. In particular, the reactive response of such a strong $\chi^{(3)}$ material would lead to distortions of the temporal mode shapes \cite{Shapiro_2006}, departing from our computational mode basis. Different remedies have been proposed to circumvent this issue, such as temporal mode trapping~\cite{Yanagimoto_Ng_Jankowski_Mabuchi_Hamerly_2022}, however significantly increasing the complexity of the simulator. Although bringing this platform towards the fermionized limit constitutes an outstanding technical challenge, we emphasize that the proposed TMN architecture makes optimal use of the nonlinear resource by recycling it for every site at every round-trip. To date, many routes have been shown to provide such strong non-linearities, such as trapped atoms and ions, color centers, and quantum dots \cite{Chang_Vuletić_Lukin_2014, Kala_Sharp_Choi_Manna_Deshmukh_Kizhake_Veetil_Menon_Pelton_Waks_Majumdar_2025}. In this context, dynamical coupling schemes for gating the light in and out of resonators~\cite{heuck2020photon, Karni_Vaswani_Chervy_2025} offers a promising route to coherently swap pulses between our time-multiplexed simulator and an auxiliary nonlinear cavity.  Finally, we note that the present architecture could be revisited in the context of superconducting microwave circuits where sources of strong nonlinearities are readily available.

\section{conclusion}
\label{Sec:Conclusion}

In this work, we have proposed a time-multiplexed photonic network architecture capable of emulating, in the Suzuki–Trotter limit, the dynamics of Hopfield-like Hamiltonians over a wide range of parameters, including system size, site detunings, energetic disorder, and nonlinear interactions. The platform relies on discrete-time coupling events between temporal modes of light in coupled ring resonators, yielding an effective stroboscopic evolution that accurately reproduces the continuous-time dynamics of the bosonized Hopfield model.

We have shown that the proposed architecture is compatible with existing photonic technologies, such as integrated photonic circuits and fiber-based telecom networks, and therefore can be scaled to large effective system sizes. The time-multiplexed nature of the simulator enables precise and independent control over site energies and coupling conditions through dynamical modulation, offering practical advantages over spatially multiplexed platforms. In particular, it opens numerous possibilities for the study and control of sub-Rabi-cycle dynamics, non-adiabatic quenches, and individual stochastic trajectories in driven-dissipative systems. By benchmarking the stroboscopic evolution against the corresponding continuous-time dynamics, we find that effective system sizes of order 1000 sites should be achievable with currently available fabrication and modulation capabilities. Moreover, dynamical coupling schemes~\cite{heuck2020controlled} and alternative network architectures~\cite{bartlett2024programmable} could enable the emulation of transport properties in strongly coupled light-matter systems~\cite{hagenmuller2018cavity, dubail2022large}.

In addition to linear dynamics, we have analyzed the behavior of the simulator in the presence of Kerr-type self-phase modulation. In the weakly nonlinear regime, the network reproduces the expected mean-field bistable response, in agreement with established results. For stronger nonlinearities, our analysis indicates that the platform may be extended to emulate nonlinear Hamiltonian dynamics beyond the mean-field description, providing a potential route toward the optical simulation of interacting light–matter models such as the Tavis–Cummings Hamiltonian. More generally, our results place time-multiplexed resonator as a viable framework for analog simulation of strongly coupled light-matter systems.

\bigbreak

\newpage

\begin{acknowledgments}
 We are thankful to Sara Bassil for her insights about Si-integrated photonics and Maxime Dherbécourt for the fruitful discussions on Hopfield-like models.
\end{acknowledgments}

\appendix

\bibliography{References.bib}

\end{document}